\documentstyle[11pt,newpasp,twoside,epsf]{article}
\def\edcomment#1{\iffalse\marginpar{\raggedright\sl#1\/}\else\relax\fi}
\reversemarginpar

\begin{document}
\title{$\tau$ Boo b: Not so bright, but just as heavy}
 \author{Andrew Collier Cameron, Keith Horne, David James}
\affil{School of Physics and Astronomy,
University of St Andrews,
North Haugh,
St Andrews, Fife,
SCOTLAND KY16 9SS}
\author{Alan Penny}
\affil{Rutherford Appleton Laboratory, Chilton, Didcot, Oxon ENGLAND OX11 0QX}
\author{Meir Semel}
\affil{DASOP, Observatoire de Paris-Meudon, F-92195 Meudon-Cedex, France}

\begin{abstract}
We present new results derived from high-resolution optical spectra of
the $\tau$ Boo system, secured in March-May 2000.  The results do not
show the same feature reported by Cameron et al (1999) as a candidate
reflected-light signature from the planet.  Together
with earlier results from the 1998 and 1999 seasons, the new
data yield a 99.9\%\
upper limit on the opposition planet/star flux ratio $\epsilon < 
3.5\times 10^{-5}$ between 387.4 and
586.3 nm, a factor 3 deeper than the upper limit of 
Charbonneau et al (1999). For an assumed planet radius $R_{p}= 1.2 R_{J}$, 
the upper limit on the mean geometric albedo is $p < 
0.22$, 40\%\ that of Jupiter.
We find new evidence that the star's rotation is
synchronised with the planet's orbital motion.  Using a Monte Carlo
analysis we infer that the planet's mass must lie in the range 5.5 to
10 times the mass of Jupiter.
\end{abstract}

\section{Introduction and observations}

During the last 5 years, many Jupiter-mass planets have been
discovered in 3 to 5-day orbits about solar-type stars.  Determining
the atmospheric composition of such planets presents challenges to
both observers and theorists.  Theoretical models must address the
equilibrium chemistry, cloud-formation physics and
absorption/scattering properties of irradiated atmospheres at
temperatures of order 1000 K. Several such models (Marley et al 1999;
Sudarsky et al 2000; Seager \&\ Sasselov 2000; Seager et al 2000) have
been published lately, with a variety of predictions for the geometric
albedo spectra.  For observers, the challenge is to disentangle the
optical reflection spectrum or the thermal IR emission spectrum from a
stellar background that may be tens of thousands of times brighter. 
Charbonneau et al (1999) published a deep upper limit on the strength
of the mid-optical reflection of the planet that orbits t Boo every
3.3 days, while Cameron et al (1999) reported a candidate
reflected-light signature at flux levels slightly greater than the
Charbonneau upper limit.

We secured new observations of $\tau$ Boo on 2000 March 14, 15, 24,
April 23, 24, May 13 and 17.  We used the same instrument and detector
on the 4.2-m William Herschel Telescope, as previously reported by
Collier Cameron et al (1999; CHPJ99) for the 1998 and 1999 data.  The
observational procedures and data extraction and analysis methods were
identical to those described by CHPJ99.  An updated orbital ephemeris
for the times at which the planet's velocity passes through the
centre-of-mass velocity from red to blue, HJD$ = 2451652.312 +
3.312450E$, was derived from radial-velocity data kindly provided by
G. Marcy and colleagues, incorporating observations as recent as 2000
March 24.  This leads to a small but significant departure from the
extrapolated orbital timings used in CHPJ99's interpretation of the
1998/99 data.  

\section{Deep upper limit on albedo}

Analysis of all three seasons' WHT data with the revised ephemeris
does not reproduce the candidate detection reported by CHPJ99. 
Instead we find a 99.9\%\ upper limit on the opposition planet/star
flux ratio, $\log\epsilon < -4.45$, in the most likely range of orbital
velocity amplitudes, $70 < K_{p} < 100$ km s$^{-1}$ (Fig.  1).  This
value for $\epsilon$ assumes a wavelength-independent albedo over the
observed wavelength range 387.4 to 586.3 nm, and a Lambert-sphere
phase function.  Given that $\epsilon=p(R_{p}/a)^{2}$ where $a$ is the
radius of the planet's orbit, the geometric albedo must be $p<0.22$ if
$R_{p}\simeq 1.2 R_{J}$ as indicated by the recent models of Burrows
et al (2000).

\begin{figure}[t]
\plotfiddle{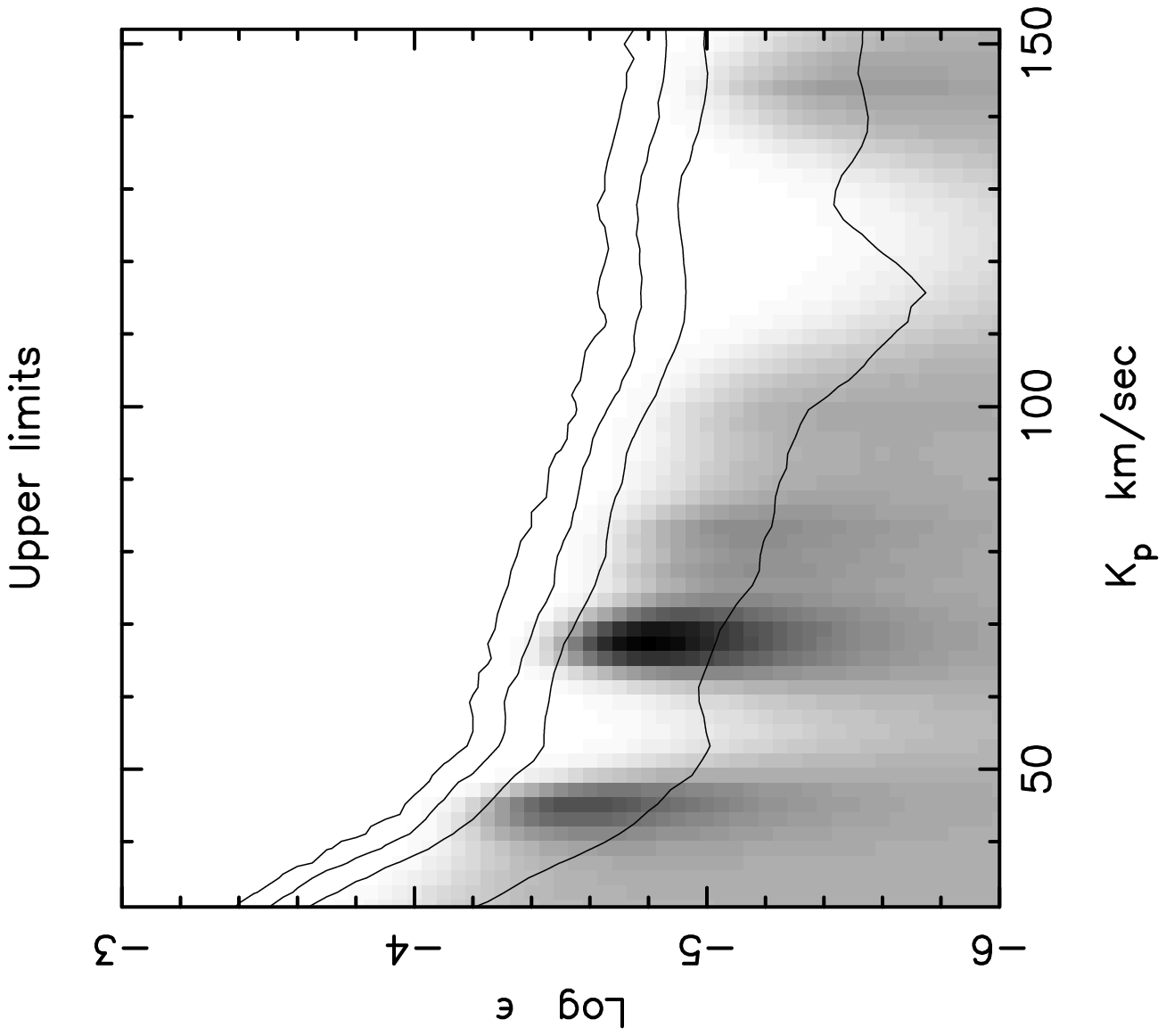}{2.0in}{270}{50}{50}{-210pt}{205pt}
\plotfiddle{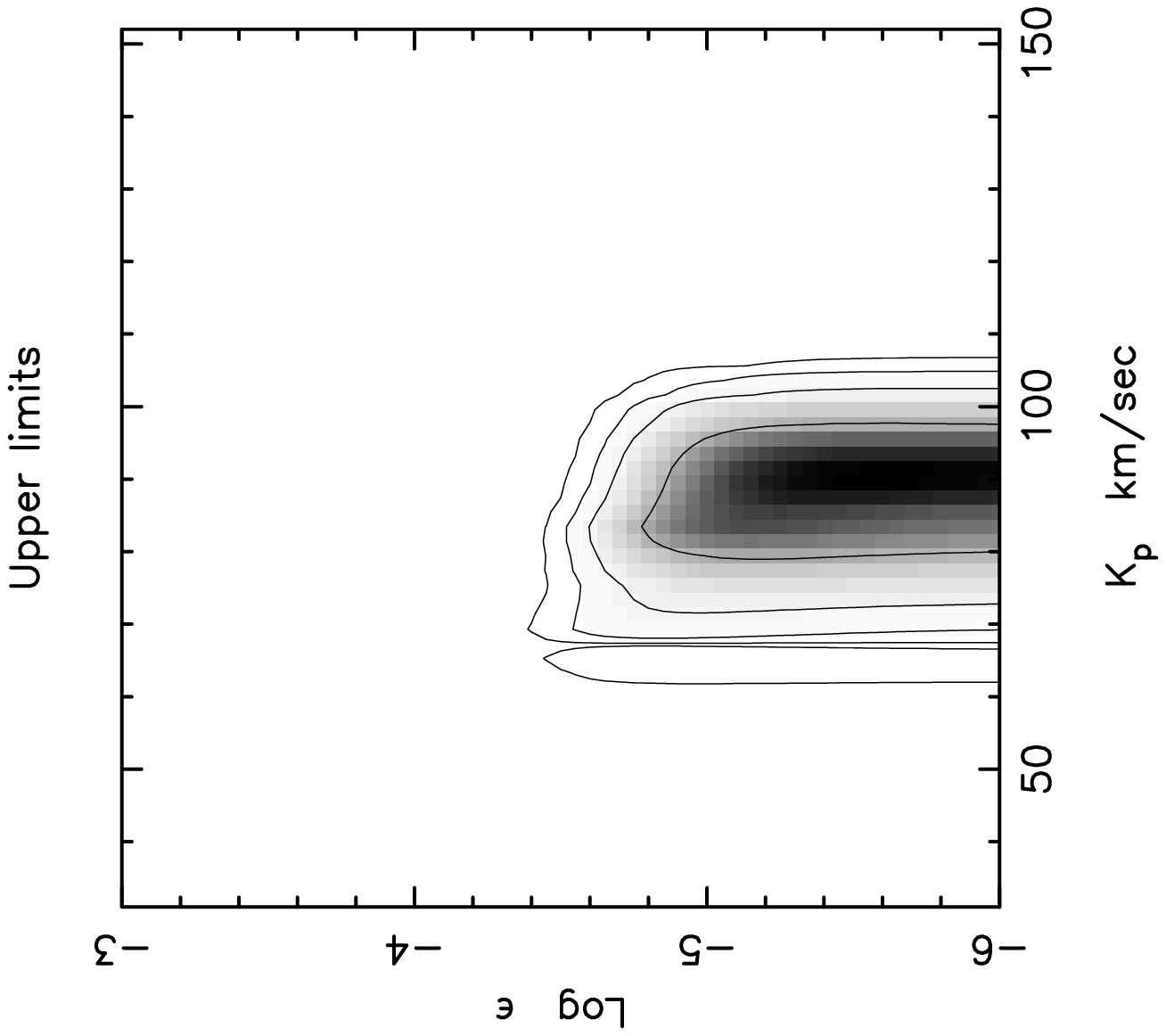}{0.0in}{270}{50}{50}{-30pt}{230pt}
\caption[]{The greyscale gives the relative probability of the fit to
the data as a function of opposition planet/star flux ratio $\epsilon$
and projected orbital velocity amplitude $K_{p}$.  At left, no
constraints are placed on the orbital inclination.  At right, the
probabilities are modified by the prior probability distribution for
Kp assuming the star's rotation is synchronous (see Section 3).  The
contours enclose 68.3, 95.4, 99.0 and 99.9 percent of the
probability.}
\label{fig:epskp}
\end{figure}

\section{Synchronous rotation and orbit inclination}

The HIPPARCOS parallax and Barnes-Evans angular diameter yield
$R_{*}=1.5\pm 0.2$ R$_{\odot}$ (Baliunas et al 1995).  The spectral
type and surface gravity indicate $M_{*}=1.4 \pm 0.1$ M$_{\odot}$
(Fuhrmann 1998, Gonzalez 1998).  The orbital radial velocity amplitude
of the star is $K_{*} = 466.0 \pm 2.3$ m s$^{-1}$ (Marcy 1999,
personal communication).  The projected equatorial rotation speed of
the star is $v \sin i = 14.8 \pm 0.5$ km s$^{-1}$ (Gray 1982, Baliunas
et al 1997, Fuhrmann 1998, Gonzalez 1998, Cameron et al 1999).  We
performed Monte Carlo simulations with random Gaussian distributions
in these 4 variables to determine the distributions for derived planet
mass and projected orbital velocity amplitude.

At the $>1$ Gyr age of $\tau$ Boo, the high $v\sin i$ suggests tidal
synchronisation.  As in 1998 and 1999, the spring 2000
data show distortions in the stellar line profiles, which drift from
blue to red at the rate expected of stellar surface features if the
star rotates synchronously with the planet's orbit.  We therefore
rejected models that yielded synchronization timescales for the star,
$$
\tau_{sync}\simeq 1.2 
\left(\frac{M_{p}}{M_{*}}\right)^{-2}\left(\frac{a}{R_{*}}\right)^{6}\mbox{ years,}
$$ 
that were longer than the main-sequence lifetime $\tau_{ms}\simeq
10^{10}(M_{*}/M_{\odot})^{-3}$ years.  The resulting probability
distribution spans the orbital velocity range $70 < K_{p}< 110$ km
s$^{-1}$ and yields planet masses $5.5 < M_{p}/M_{J} < 10$.

The new upper limit on the geometric albedo of $\tau$ Boo b, $p<0.22$
between 387.4 to 586.3 nm, lies between the albedo spectra predicted
by Sudarsky et al (2000) for ``isolated'' and ``modified'' Class IV
roaster atmospheres, and suggests that substantial pressure-broadened
Na I D absorption may be present.  The Class V models of Sudarsky et
al appear to offer better prospects for direct detection of hot
Jupiters with lower surface gravities such as $\upsilon$ And b, for
which similar observations are scheduled in 2000 October and November.

\end{document}